\documentclass[twocolumn,showkeys,aps,prb,showpacs]{revtex4-1}
\usepackage{graphicx}
\usepackage[CJKbookmarks,dvipdfm,colorlinks,linkcolor=blue,citecolor=blue]{hyperref}

\begin{document}

\title{Huge out-of-plane piezoelectric response in ferromagnetic monolayer NiClI}

\author{San-Dong Guo$^{1}$, Yu-Tong Zhu$^{1}$,  Ke Qin$^{1}$ and Yee-Sin Ang$^{2}$}
\affiliation{$^1$School of Electronic Engineering, Xi'an University of Posts and Telecommunications, Xi'an 710121, China}
\affiliation{$^2$Science, Mathematics and Technology (SMT), Singapore University of Technology and Design (SUTD), 8 Somapah Road, Singapore 487372, Singapore}
\begin{abstract}
 The combination of piezoelectricity and ferromagnetic (FM) order in a two-dimensional (2D) material, namely 2D piezoelectric ferromagnetism (PFM), may open up unprecedented opportunities for novel device applications.  Here,  we predict an  in-plane FM semiconductor Janus monolayer NiClI with considerably large  magnetic
anisotropy energy (MAE) of 1.439 meV,  exhibiting  dynamic, mechanical and thermal  stabilities. The NiClI monolayer
possesses larger  in-plane piezoelectricity ($d_{11}$$=$5.21 pm/V) comparable to that of $\mathrm{MoS_2}$. Furthermore, NiClI has huge out-of-plane piezoelectricity ($d_{31}$$=$1.89 pm/V), which is highly desirable for ultrathin piezoelectric device application. It is proved that  huge out-of-plane piezoelectricity is robust
against electronic correlation, which confirms reliability of huge $d_{31}$. Finally, being analogous to NiClI, PFM with large out-of-plane piezoelectricity can also be achieved in the Janus  monolayers of  NiClBr and NiBrI, with the predicted $d_{31}$ of 0.73 pm/V and  1.15 pm/V, respectively.  The predicted huge out-of-plane piezoelectric response  makes Janus monolayer NiClI  a good platform for multifunctional semiconductor spintronic applications, which is also compatible with the
 bottom/top gate technologies of conventional semiconductor nanoelectronic devices.

\end{abstract}
\keywords{Ferromagnetism, Piezoelectronics, 2D materials ~~~~~~~~~~~~~~~~~~~~~~~~~~~~~Email:sandongyuwang@163.com}

\maketitle

\section{Introduction}
Two-dimensional (2D) materials can exhibit significantly different and novel properties
compared to their bulk counterparts\cite{q1,q2,q4-1,q4}.  A particularly useful property is piezoelectricity, which can convert mechanical energy into electrical
energy and vice versa\cite{q4-1,q4}. The piezoelectricity of 2D materials has
been widely investigated in recent years.
Experimentally, the piezoelectricity  has been confirmed in multiple 2D materials, such as $\mathrm{In_2Se_3}$\cite{q8-1},  $\mathrm{MoS_2}$\cite{q5,q6} and  MoSSe\cite{q8},  thus concretely establishing the feasibility of 2D-material-based piezoelectric device technology. Theoretically,  a large number of 2D
materials have been predicted to be piezoelectric, including transition metal dichalchogenides (TMD), Janus TMD, $\mathrm{MA_2Z_4}$, group IIA and IIB metal oxides, group-V binary semiconductors and group III-V semiconductors\cite{q7,q7-1,q7-2,q7-3,q7-4,q7-5, q7-6,q7-7}.
A substantial in-plane piezoelectricity can be observed in many 2D materials\cite{q7,q7-2,q7-3,q7-4,q7-5}, while an  out-of-plane piezoelectric response is typically very small\cite{q7,q7-2} (For most of them, the $d_{31}$  is less than 1 pm/V).  Thus,  searching for large out-of-plane piezoelectric response in 2D materials has become an important quest, particularly due to its compatibility with bottom/top gate device architecture of many conventional semiconductor devices.

The multifunctional 2D piezoelectric  materials, such as coexistence of piezoelectricity and  electronic topology or/and ferromagnetism or/and valley,  are of particular interest, which can provide a potential platform for multi-functional electronic devices. The PFMs have been predicted in 2D vanadium dichalcogenides,  $\mathrm{VSi_2P_4}$, $\mathrm{CrBr_{1.5}I_{1.5}}$ and $\mathrm{InCrTe_3}$\cite{qt1,q15,q15-1,q15-2}. The  piezoelectric quantum anomalous Hall insulator (PQAHI), which  combines piezoelectricity with topological and FM orders, has been reported in Janus monolayer  $\mathrm{Fe_2IX}$ (X=Cl and Br)\cite{gsd1}. This provides possibility to use the piezoelectric effect to control quantum anomalous Hall (QAH) effects.
The piezoelectric ferrovalley (FV) material  has been proposed in $\mathrm{GdCl_2}$ monolayer\cite{gsd2}, and the anomalous valley Hall effect may be achieved  by piezoelectric effect. Therefore, the search of high-performance 2D piezoelectric materials is a critical step that may open new opportunities towards the development of novel nanodevice applications.
\begin{figure*}
  \includegraphics[width=12cm]{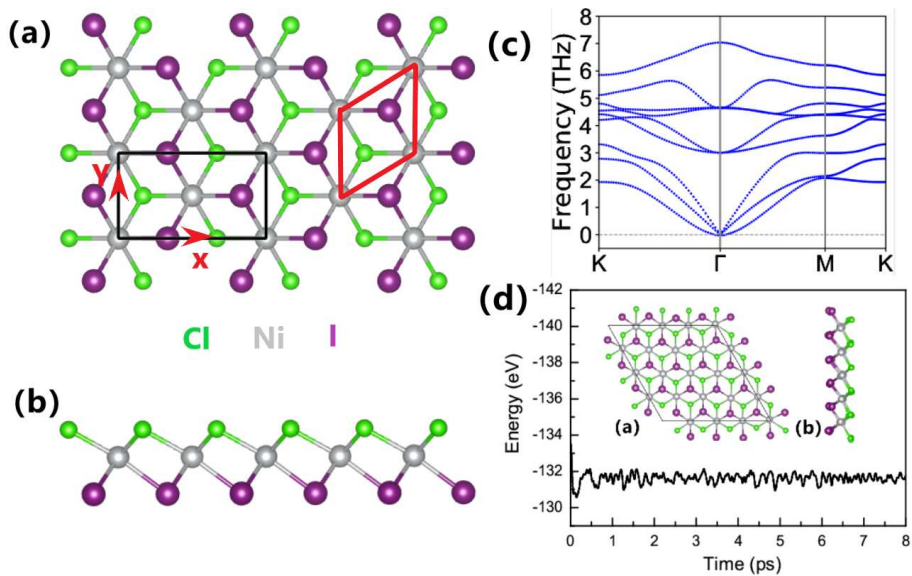}
  \caption{(Color online)For Janus NiClI monolayer,  (a): top view and (b): side view of  crystal structure. The primitive (rectangle supercell) cell is
   marked by red (black) lines. (c):the phonon dispersion curves. (d):the variation of free energy during the 8 ps AIMD simulation. Insets show the
 final structures (top view (a) and side view (b)) after 8 ps at 300 K.}\label{t0}
\end{figure*}

Multifunctional 2D piezoelectric  materials with large out-of-plane piezoelectric response has received strong research attention in recent years. Magnetic monolayer metal halides $\mathrm{MX_2}$ (M=V, Cr, Mn, Fe, Co and Ni; X=Cl, Br and I) with 1T crystal
structure ($C_{3v}$ symmetry) have been investigated from first-principles and Monte Carlo simulations, and $\mathrm{NiX_2}$ (X=Cl, Br and I) of them are FM semiconductors with excellent stability\cite{e,e1}.
Due to the inversion symmetry, $\mathrm{NiX_2}$ (X=Cl, Br and I) monolayers are not  piezoelectric.  Janus 2D materials with breaking mirror or inversion
symmetry along out-of-plane orientations provide  possibility to achieve out-of-plane piezoelectric response. Based on  $\mathrm{MoS_2}$, Janus monolayer MoSSe has been successfully synthesised  by various
experimental synthesis strategies\cite{q8,e2}.

Motivated by the successful demonstration of strong out-of-plane piezoelectric response in Janus system,   we  explore the electronic and piezoelectric properties of
 Janus NiClI monolayer, which can be constructed from $\mathrm{NiCl_2}$/$\mathrm{NiI_2}$ monolayer. It is found that NiClI monolayer is an  in-plane FM semiconductor with large MAE of 1.439 meV,  exhibiting good  stability. The NiClI monolayer
not only possesses large in-plane piezoelectricity ($d_{11}$$=$5.21 pm/V), but also has huge out-of-plane piezoelectricity ($d_{31}$$=$1.89 pm/V). Calculated results show that the excellent piezoelectricity is robust against electronic correlation. Finally, similar to NiClI, the PFMs with large out-of-plane piezoelectric response can also be realized  in the  monolayer NiClBr and NiBrI. These findings reveal   the enormous potential of NiClI monolayer
as a promising candidate materials for the development of novel 2D piezoelectric spintronic  devices.

\begin{figure}
  \includegraphics[width=8cm]{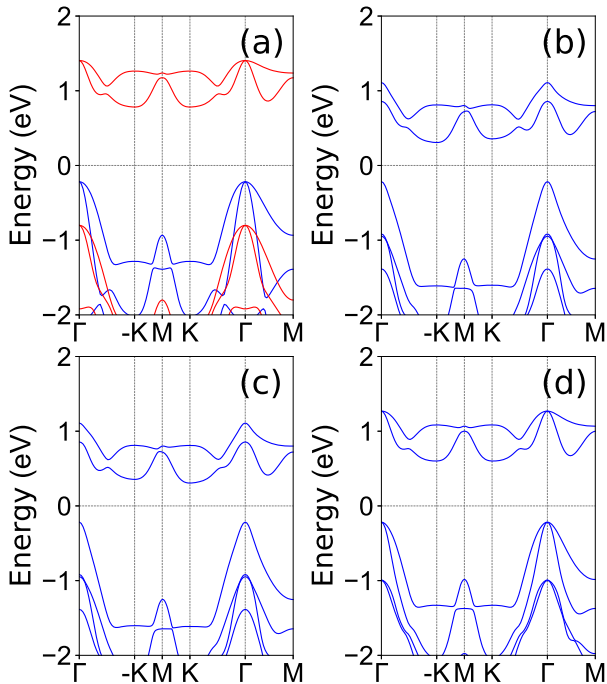}
  \caption{(Color online)The energy  band structures of  monolayer NiClI (a) without SOC; (b), (c) and (d) with SOC for magnetic moment of Ni along the positive $z$, negative $z$, and positive $x$ direction, respectively.  In (a), the blue (red) lines represent the band structure in the spin-up (spin-down) direction.}\label{band}
\end{figure}

\begin{table*}
\centering \caption{For monolayer  NiXY (X/Y=Cl, Br and I; X$\neq$Y), the lattice constants $a_0$ ($\mathrm{{\AA}}$), the energy difference between AFM and FM ordering $\Delta E$ (meV), the magnetic anisotropy energy MAE ($\mathrm{\mu eV}$),   the elastic constants $C_{ij}$ ($\mathrm{Nm^{-1}}$), the  piezoelectric coefficients   $e_{ij}$  ($10^{-10}$ C/m ) and  $d_{ij}$ (pm/V). }\label{tab0}
  \begin{tabular*}{0.96\textwidth}{@{\extracolsep{\fill}}cccccccccc}
  \hline\hline
 Name &$a_0$&$\Delta E$ & MAE& $C_{11}$&$C_{12}$&$e_{11}$&$e_{31}$&$d_{11}$&$d_{31}$\\\hline\hline
$\mathrm{NiClBr}$ &     3.604  &   23.28    &  -45    & 47.61   & 13.55      & 0.46& 0.45   &   1.35   &  0.73                           \\\hline
$\mathrm{NiBrI}$ &     3.855  &   24.91   &  -749     & 41.95    & 12.21    & 0.74  & 0.62   &  2.49    & 1.15                       \\\hline
$\mathrm{NiClI}$ &     3.774   &   18.84   &  -1439     & 42.33    & 13.41    & 1.51  & 1.05   &   5.21  & 1.89
\\\hline\hline
\end{tabular*}
\end{table*}

\section{Computational detail}
The density functional theory (DFT)  calculations\cite{1} are performed by using the projector-augmented wave method as implemented
in the plane-wave code VASP\cite{pv1,pv2,pv3}. The popular generalized gradient approximation  of Perdew, Burke and  Ernzerhof  (GGA-PBE)\cite{pbe}  is adopted as the exchange-correlation  functional.  To consider  on-site Coulomb correlation of  Ni-3$d$ electrons,
the GGA+$U$ scheme by the
rotationally invariant approach proposed by Dudarev et al\cite{u} is adopted with $U$$=$4 eV, where only the effective
$U$ ($U_{eff}$) based on the difference between the on-site Coulomb interaction
parameter  and exchange parameters  is meaningful.   The spin-orbital coupling (SOC) effect is explicitly included in the calculations of  MAE and electronic structures.
The energy cut-off of 500 eV, total energy  convergence criterion of  $10^{-8}$ eV and force
convergence criteria  of less than 0.0001 $\mathrm{eV.{\AA}^{-1}}$   on each atom  are adopted to attain reliable results.  A vacuum space of more than 18 $\mathrm{{\AA}}$
along the $z$ direction is adopted to avoid interactions between
two neighboring images.
The phonon dispersions  are calculated
 through the direct supercell method with the 5$\times$5$\times$1 supercell, as implemented in Phonopy code\cite{pv5}.
The coefficients of the elastic  stiffness/piezoelectric stress  tensors  $C_{ij}$/$e_{ij}$   are calculated by using strain-stress relationship (SSR)/density functional perturbation theory (DFPT) method\cite{pv6}.
 The Brillouin zone sampling
is done using a Monkhorst-Pack mesh of 21$\times$21$\times$1  for $C_{ij}$, and  12$\times$21$\times$1 for $e_{ij}$.
The 2D elastic coefficients $C^{2D}_{ij}$
 and   piezoelectric stress coefficients $e^{2D}_{ij}$
have been renormalized by the the length of unit cell along $z$ direction ($L_z$):  $C^{2D}_{ij}$=$L_z$$C^{3D}_{ij}$ and $e^{2D}_{ij}$=$L_z$$e^{3D}_{ij}$.

\section{Main calculated results}
As shown in \autoref{t0} (a) and (b), the crystal structure of NiClI is formed by a Cl-Ni-I sandwich layer, and the Ni atom
is surrounded by a  distorted octahedron composed of three Cl and three I atoms.  The NiClI monolayer can be constructed   by replacing one of two  Cl/I   layers with I/Cl atoms in $\mathrm{NiCl_2}$/$\mathrm{NiI_2}$. Due to the broken central
symmetry, the symmetry of NiClI
monolayer (No.156) is lower than that of $\mathrm{NiCl_2}$/$\mathrm{NiI_2}$  monolayer (No.164).  Due to different electronegativities between Cl and I atoms, a natural vertical built-in electric field appears in the intralayer, which induces an out-of-plane piezoelectric response.
The FM and antiferromagnetic (AFM) magnetic configurations are considered to
ascertain the magnetic ground state of NiClI. The energy difference between AFM and FM orderings is 18.84 meV, indicating that  the ground state of NiClI is FM.
The optimized lattice constant $a$ is  3.774 $\mathrm{{\AA}}$, which is between ones of $\mathrm{NiCl_2}$ (3.518 $\mathrm{{\AA}}$) and $\mathrm{NiI_2}$ (3.983 $\mathrm{{\AA}}$)\cite{e1}.

As shown in \autoref{t0} (c), the phonon spectrum of NiClI shows no imaginary frequencies, indicating its dynamic stability.
Two in-plane acoustic branches display linear dispersions, while  out-of-plane acoustic branch shows a quadratic dispersion.  These share  general characteristics of 2D materials, when they are free of stress\cite{r1,r2}.
To confirm the thermal stability, the evolution of total
energy of NiClI as a function of simulation time  is calculated using ab initio
molecular dynamics (AIMD) with  a 4$\times$4$\times$1
supercell   in the canonical ensemble for 8 ps with a
time step of 1.0 fs, which is plotted in \autoref{t0} (d). During the simulation period, the frameworks of NiClI are well preserved with  little energy fluctuation, confirming its thermal stability.  The elastic constants ( $C_{11}$=42.33 $\mathrm{Nm^{-1}}$ and $C_{12}$=13.41 $\mathrm{Nm^{-1}}$) of NiClI meet
 Born  criteria of  mechanical stability of a material with hexagonal symmetry ($C_{11}>0$ and  $C_{11}-C_{12}>0$)\cite{ela},  indicating its mechanical stability.

 We show the spin-polarized energy band structures of NiClI  in \autoref{band} by using GGA and GGA+SOC.
 \autoref{band} (a) shows  a distinct
spin splitting  due to the exchange
interaction, and  NiClI is  an indirect band
gap semiconductor  with gap of 1.00 eV. The valence band maximum (VBM)/conduction band bottom (CBM) is  provided by the spin-up/spin-down, which is at $\Gamma$/K (-K) point. It is noted that the energies of  -K and K valleys are degenerate.
As plotted in FIG.1 of electronic supplementary information (ESI), the Ni-$d$ projected band structures show  that $d_{x^2-y^2}$/$d_{xy}$ orbitals  dominate -K and K valleys of conduction band, which is very key to produce valley polarization. When the SOC is included, the spontaneous valley polarization can be observed, and  the valley splitting of bottom conduction band   is 49  meV. The  K valley has  higher energy than -K valley, and  the valley polarization can  be
switched (The energy of -K valley
is higher than one of K valley.) by reversing the magnetization direction, as shown in \autoref{band} (c).
It should be pointed out that $\mathrm{NiCl_2}$/$\mathrm{NiI_2}$  monolayer has no spontaneous valley polarization due to possessing centrosymmetry.
\autoref{band} (b) and (c) show that NiClI is still   an indirect band
gap semiconductor with gap value of 0.523 eV. As shown in \autoref{band} (d),  when the  magnetization direction of NiClI is adjusted to positive $x$ direction, no valley polarization can be observed, and an indirect gap of 0.818 eV can be observed.

If the  $d_{x^2-y^2}$/$d_{xy}$ orbitals dominate  -K and K valleys, the valley splitting $|\Delta E|$ between  -K and K points  can be expressed as\cite{v2,v3}:
\begin{equation}\label{m4}
|\Delta E|=E^{K}-E^{-K}=4\alpha
\end{equation}
where $\alpha$ is the strength of SOC. If the -K and K valleys are mainly from $d_{z^2}$ orbitals, the valley splitting $|\Delta E|$
 is written as:
\begin{equation}\label{m4}
|\Delta E|=E^{K}-E^{-K}=0
\end{equation}
When the $d_{x^2-y^2}$/$d_{xy}$ orbitals dominate -K/K valley with the magnetization orientation being general case, $\Delta E=4\alpha cos\theta$\cite{v3} ($\theta$=0/90$^{\circ}$ means out-of-plane/in-plane direction.). For in-plane  magnetocrystalline direction, the valley splitting of NIClI monolayer will disappear.

  \begin{figure}
  \includegraphics[width=8cm]{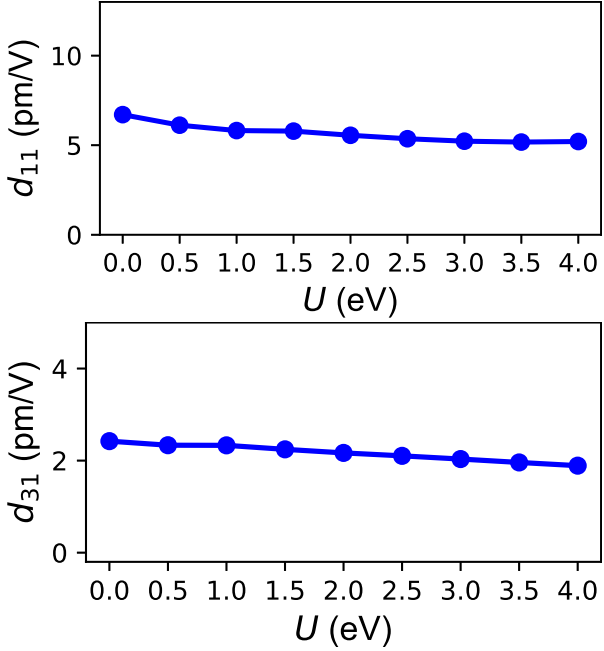}
  \caption{(Color online) For monolayer NiClI, the piezoelectric strain coefficients  $d_{11}$/$d_{31}$   as a function of  $U$.  }\label{ed1}
\end{figure}

Although the magnetocrystalline direction  of
NiClI monolayer can be regulated by external magnetic field, the intrinsic magnetic anisotropy should be determined. The MAE defined as  energy difference
($E_x$-$E_z$) is calculated, where $E_x$/$E_z$ is the energy per Ni atom when the magnetization is along the $x$/$z$ direction.
The positive value means out-of-plane direction, while the negative value suggests  in-plane one. By using GGA+SOC, the calculated MAE is -1439 $\mathrm{\mu eV}$, which means that the intrinsic easy axis of NiClI is in-plane. So, NiClI monolayer is a common FM semiconductor, not a FV material.

The NiClI monolayer shows both in-plane and  out-of-plane piezoelectric response due to special Janus structure. By performing symmetry analysis,  only considering  in-plane strain and stress, the  piezoelectric stress,   piezoelectric strain and elastic tensors of NiClI by using  Voigt notation can be written as\cite{q7,q7-2}:
 \begin{equation}\label{pe1-1}
 e=\left(
    \begin{array}{ccc}
      e_{11} & -e_{11} & 0 \\
     0 & 0 & -e_{11} \\
      e_{31} & e_{31} & 0 \\
    \end{array}
  \right)
    \end{equation}

  \begin{equation}\label{pe1-2}
  d= \left(
    \begin{array}{ccc}
      d_{11} & -d_{11} & 0 \\
      0 & 0 & -2d_{11} \\
      d_{31} & d_{31} &0 \\
    \end{array}
  \right)
\end{equation}
\begin{equation}\label{pe1-4}
   C=\left(
    \begin{array}{ccc}
      C_{11} & C_{12} & 0 \\
     C_{12} & C_{11} &0 \\
      0 & 0 & (C_{11}-C_{12})/2 \\
    \end{array}
  \right)
\end{equation}
   With an imposed uniaxial in-plane strain,   both $e_{11}$/$d_{11}$$\neq$0 and $e_{31}$/$d_{31}$$\neq$0. However, when   a biaxial in-plane strain is applied,
$e_{11}$/$d_{11}$=0, but $e_{31}$/$d_{31}$$\neq$0.
Here, the two independent $d_{11}$ and $d_{31}$ can be derived by $e_{ik}=d_{ij}C_{jk}$:
\begin{equation}\label{pe2}
    d_{11}=\frac{e_{11}}{C_{11}-C_{12}}~~~and~~~d_{31}=\frac{e_{31}}{C_{11}+C_{12}}
\end{equation}

The orthorhombic supercell (see  \autoref{t0}) is adopted  to calculate the  $e_{11}$/$e_{31}$ of NiClI.
By DFPT method, the attained $e_{11}$/$e_{31}$ is 1.51$\times$$10^{-10}$/1.05$\times$$10^{-10}$ C/m  with ionic part 1.35$\times$$10^{-10}$/-0.03$\times$$10^{-10}$ C/m  and electronic part 0.16$\times$$10^{-10}$/1.08$\times$$10^{-10}$ C/m. For $e_{11}$, the electronic and ionic contributions have  the same signs, and   the ionic part
 dominates the  piezoelectricity. However, for $e_{31}$, the electronic and ionic contributions  have  opposite signs, and   the electronic part
 plays a decisive role.

  Based on \autoref{pe2}, the corresponding  $d_{11}$/$d_{31}$ of NiClI monolayer is  5.21/1.89 pm/V.
To be compatible with the conventional bottom/top gate technologies,  a large out-of-plane piezoelectric response is
highly desired for 2D materials.  The predicted  $d_{31}$ of NiClI monolayer is  higher than many  2D known materials, including the oxygen functionalized MXenes (0.40-0.78 pm/V)\cite{q9},  $\mathrm{Sb_2Te_2Se}$ (1.72 pm/V)\cite{re-3},  Janus TMD monolayers (0.03 pm/V)\cite{q7},
functionalized h-BN (0.13 pm/V)\cite{o1}, Janus $\mathrm{CrCl_{1.5}I_{1.5}}$/$\mathrm{CrBr_{1.5}I_{1.5}}$ monolayer (1.14-1.80 pm/V)\cite{q15-1}, kalium decorated graphene (0.3
pm/V)\cite{o2}, Janus group-III materials (0.46 pm/V)\cite{q7-6-1}, Janus BiTeI/SbTeI  monolayer (0.37-0.66 pm/V)\cite{o3}, $\alpha$-$\mathrm{In_2Se_3}$
(0.415 pm/V)\cite{o4} and MoSO (0.7 pm/V)\cite{re-11}.
The $d_{31}$ of NiClI   is lower than ones of  TePtS/TePtSe (2.4-2.9 pm/V)\cite{re-6} and $\mathrm{CrF_{1.5}I_{1.5}}$ (2.58 pm/V)\cite{q15-1}. However, compared with  TePtS/TePtSe,  the coexistence of intrinsic piezoelectricity and ferromagnetism can be achieved in NiClI.  With respect to $\mathrm{CrF_{1.5}I_{1.5}}$, the NiClI has more higher $d_{11}$. So, NiClI is a excellent PFM with huge out-of-plane piezoelectric response.

Electronic correlation has important effects on the electronic states and piezoelectric properties  of  2D materials\cite{gsd2,h10,h11}. To confirm the large $d_{31}$ in NiClI monolayer, electronic correlation effects on its piezoelectric properties are investigated. Firstly, we optimize  lattice constants $a$  at different $U$ (0-4 eV) (see FIG.2 of ESI). The $a$ increases with increasing $U$, and the change is very small (about 0.017 $\mathrm{{\AA}}$).  The energy difference between  AFM  and FM configurations  as a function of $U$ is plotted in FIG.2 of ESI, which shows that NiClI is always a FM ground state.
For both  out-of-plane and in-plane magnetic anisotropy, the evolutions of electronic band
structures as a function of $U$ are calculated by using GGA+SOC. The energy band structures at  representative $U$ values are plotted in FIG.3 of ESI, and the gaps vs $U$ are shown in FIG.4 of ESI. It is clearly seen that NiClI changes from metal to semiconductor for out-of-plane  magnetic anisotropy. However, NiClI is always a semiconductor for in-plane case within the considered $U$ range. To determine the intrinsic magnetic anisotropy, the MAE as a function of $U$ is plotted in FIG.5 of ESI, indicating an in-plane magnetic anisotropy within the considered $U$ range. Thus, NiClI is a FM semiconductor within the considered $U$ range, which is particularly beneficial for piezoelectric device applications where the strong suppression of leakage current is required.

The elastic constants  ($C_{11}$, $C_{12}$, $C_{11}$-$C_{12}$ and $C_{11}$+$C_{12}$) of NiClI monolayer as a function of $U$ are plotted in FIG.6 of ESI. It is found that these elastic constants have weak dependence on $U$. The piezoelectric  stress  coefficients  ($e_{11}$ and $e_{31}$) along  the ionic  and electronic contributions vs $U$ are shown in FIG.7 of ESI. With increasing $U$, both  $e_{11}$ and $e_{31}$ of NiClI decrease, which  leads to reduced  $d_{11}$ and $d_{31}$.
The piezoelectric  strain  coefficients ($d_{11}$ and $d_{31}$) of monolayer  NiClI vs $U$ are plotted  in \autoref{ed1}.
It is found that reduced $U$ is in favour of enhanced  $d_{11}$ and $d_{31}$. At $U$$=$0 eV, the $d_{11}$ and $d_{31}$ can improve to 6.71 pm/V and 2.42 pm/V.
Thus, the large $d_{31}$ in NiClI monolayer is robust against electronic correlation.

\begin{figure}
  \includegraphics[width=8cm]{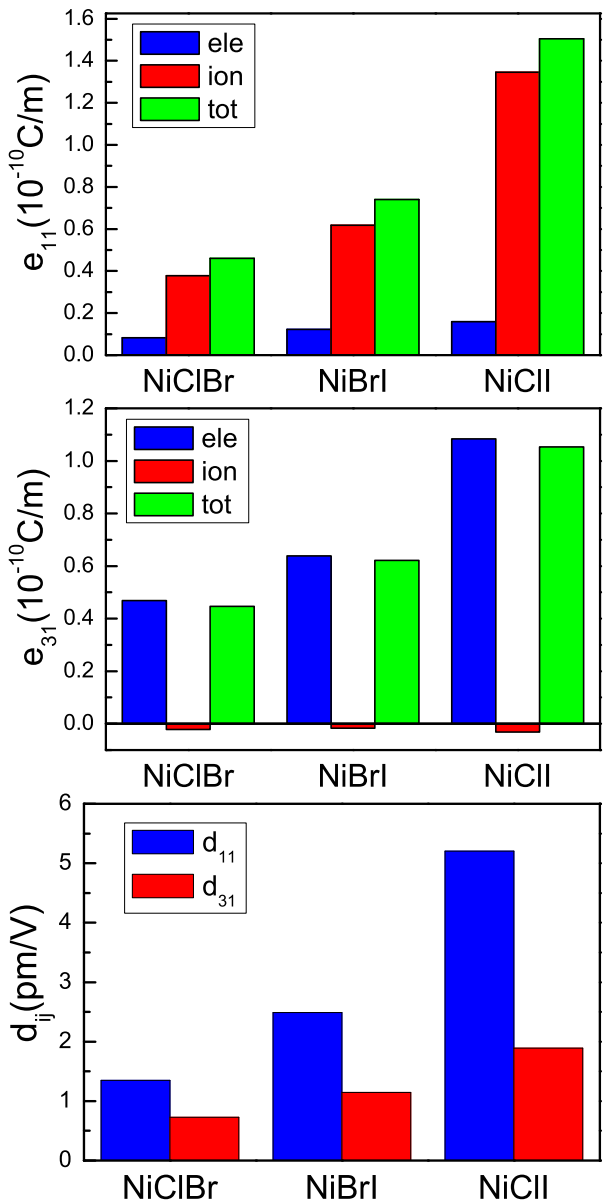}
  \caption{(Color online) For NiClBr, NiBrI and NiClI  monolayers: the piezoelectric stress coefficients  $e_{11}$/$e_{31}$ (tot) along  the ionic (ion) contribution and electronic (ele) contributions, and piezoelectric strain coefficients  $d_{11}$/$d_{31}$.}\label{ed}
\end{figure}
\section{Discussions and Conclusion}
Janus monolayer NiClBr and NiBrI can also be constructed from  $\mathrm{NiX_2}$ (X=Cl, Br and I), which may possess large $d_{31}$.
According to calculated results in \autoref{tab0}, the FM order is still the ground state of NiClBr and NiBrI by comparing the energy difference  of  AFM and FM magnetic configurations, and they  possess in-plane magnetic anisotropy based on MAE. For monolayer NiClBr and NiBrI ,  the optimized lattice constants are 3.604 $\mathrm{{\AA}}$ and 3.855  $\mathrm{{\AA}}$.  The predicted  $C_{11}$ and $C_{12}$ are  47.61 (41.95) $\mathrm{Nm^{-1}}$ and  13.55 (12.21) $\mathrm{Nm^{-1}}$ for NiClBr (NiBrI),  satisfying   Born  criteria of mechanical stability\cite{ela}. Based on phonon dispersions in FIG.8 of ESI, it is proved  that monolayer NiClBr and NiBrI are dynamically stable. The piezoelectric properties of monolayer NiClBr and  NiBrI are investigated. The  piezoelectric  stress  coefficients  ($e_{11}$ and $e_{31}$) along  the ionic/electronic contribution and piezoelectric  strain  coefficients ($d_{11}$ and $d_{31}$) of NiClBr, NiBrI and NiClI are plotted in \autoref{ed}, and the corresponding data are summarized in \autoref{tab0}.  For all three monolayers, the electronic and ionic contributions of $e_{11}$ have the same signs, and  the ionic part  dominates the  piezoelectricity. However, the electronic and ionic parts  of $e_{31}$ have  opposite signs, and   the electronic part
 dominates the  piezoelectricity with negligible ionic part. From NiClBr to NiBrI to NiClI, both $d_{11}$ and $d_{31}$ increase  due to enhanced $e_{11}$ and $e_{31}$.
A piezoelectric material should be a semiconductor, and the energy band structures of NiClBr and  NiBrI are plotted in FIG.9 of ESI with in-plane magnetic anisotropy.
It is clearly seen that they are semiconductors.

In conclusion, based on first-principle calculations,  the electronic structures and piezoelectric
properties for Janus monolayer NiClI are investigated.  For out-of-plane magnetic anisotropy, the NiClI is a FV material. However, a common FM semiconductor is observed for in-plane magnetic anisotropy. By calculating actual MAE, the easy axis of NiClI is in-plane. The calculated results show that
 NiClI monolayer has large piezoelectric response.  Especially,  the huge out-of-plane
piezoelectricity with its $d_{31}$=1.89 pm/V is achieved, which is significantly larger than those of most known 2D materials. Therefore, NiClI monolayer is  suitable for ultrathin piezoelectric devices under the $d_{31}$ operating mode. It is found that the huge $d_{31}$ is robust against electronic correlation.
Finally, similar to NiClI,  monolayer NiClBr and NiBrI are also excellent PFMs  with large $d_{31}$ being  0.73 pm/V and 1.15 pm/V.
 Our works offer useful material design guidelines   to achieve PFMs with large out-of-plane piezoelectric response.

\begin{acknowledgments}
This work is supported by the Natural Science Foundation of Shaanxi Provincial Department of Education (19JK0809). We are grateful to the Advanced Analysis and Computation Center of China University of Mining and Technology (CUMT) for the award of CPU hours and WIEN2k/VASP software to accomplish this work.
\end{acknowledgments}


\begin{references}
\bibitem{q1}Y. Liu, Y. Huang and X. F. Duan,  Nature \textbf{567}, 323 (2019).


\bibitem{q2}K. Khan, A. K. Tareen, M. Aslam, R. Wang, Y. P. Zhang, A. Mahmood, Z. B. Ouyang, H. Zhang and Z. Y. Guo,  J. Mater. Chem. C \textbf{8}, 387 (2020).

\bibitem{q4-1}Q. Zhang, S. L. Zuo, P. Chen and C. F. Pan,  InfoMat. \textbf{3}, 987, (2021).

\bibitem{q4}W. Wu and Z. L. Wang, Nat. Rev. Mater. \textbf{1}, 16031 (2016).


\bibitem{q8-1}M. Dai, Z. Wang, F. Wang, Y. Qiu, J. Zhang, C. Y. Xu, T. Zhai, W. Cao, Y. Fu,
D. Jia, Y. Zhou, and P. A. Hu, Nano Lett. \textbf{19}, 5416 (2019).

\bibitem{q5} W. Wu, L. Wang, Y. Li, F. Zhang, L. Lin, S. Niu, D. Chenet,
X. Zhang, Y. Hao, T. F. Heinz, J. Hone and Z. L. Wang,
Nature \textbf{514}, 470 (2014).


\bibitem{q6}H. Zhu, Y. Wang, J. Xiao, M. Liu, S. Xiong, Z. J. Wong, Z. Ye,
Y. Ye, X. Yin and X. Zhang, Nat. Nanotechnol. \textbf{10},
151 (2015).

\bibitem{q8}A. Y. Lu, H. Zhu, J. Xiao, C. P. Chuu, Y. Han, M. H. Chiu,
C. C. Cheng, C. W. Yang, K. H. Wei, Y. Yang, Y. Wang,
D. Sokaras, D. Nordlund, P. Yang, D. A. Muller, M. Y. Chou,
X. Zhang and L. J. Li, Nat. Nanotechnol. \textbf{12}, 744 (2017).



\bibitem{q7}L. Dong, J. Lou and V. B. Shenoy, ACS Nano, \textbf{11},
8242 (2017).


\bibitem{q7-1}Y. Xu, Z. Q. Li, C. Y. He, J. Li, T. Ouyang, C. X. Zhang, C. Tang and
 J. X. Zhong,   Appl. Phys. Lett. \textbf{116}, 023103 (2020).



\bibitem{q7-2}M. N. Blonsky, H. L. Zhuang, A. K. Singh and R.  G. Hennig,  ACS Nano  \textbf{9},
9885 (2015).


\bibitem{q7-3}S. D. Guo, Y. T. Zhu, W. Q. Mu and W. C. Ren,  EPL \textbf{132}, 57002 (2020).


\bibitem{q7-4}R. X. Fei, We. B. Li, J. Li and L. Yang, Appl. Phys. Lett.  \textbf{107}, 173104 (2015)


\bibitem{q7-5}K. N. Duerloo, M. T. Ong and E. J. Reed, J. Phys. Chem. Lett. \textbf{3}, 2871 (2012).


\bibitem{q7-6}N. Jena, Dimple, S. D.  Behere  and A. D. Sarkar, J. Phys. Chem. C  \textbf{121}, 9181 (2017).


\bibitem{q7-7}Y. Chen,  J. Y. Liu,  J. B. Yu,  Y. G. Guo and Q. Sun, Phys. Chem. Chem. Phys.
 \textbf{21}, 1207 (2019).


\bibitem{qt1}J. H. Yang,  A. P. Wang, S. Z. Zhang, J.  Liu, Z. C. Zhong and L. Chen, Phys. Chem. Chem. Phys.,
\textbf{21}, 132 (2019).

\bibitem{q15}S. D. Guo, W. Q. Mu, Y. T. Zhu and X. Q. Chen, Phys. Chem. Chem. Phys. \textbf{22}, 28359 (2020).

\bibitem{q15-1}S. D. Guo, X. S. Guo, X. X. Cai, W. Q. Mu and  W. C. Ren, J. Appl. Phys. \textbf{129}, 214301 (2021).

\bibitem{q15-2}G. Song, D. S. Li, H. F. Zhou et al., Appl. Phys. Lett. \textbf{118}, 123102 (2021).

\bibitem{gsd1}S. D. Guo, W. Q. Mu, X. B. Xiao and B. G. Liu, Nanoscale \textbf{13}, 12956 (2021).

\bibitem{gsd2}S. D. Guo, J. X. Zhu, W. Q. Mu and B. G. Liu, Phys. Rev. B \textbf{104}, 224428 (2021).

\bibitem{e}V. V. Kulish  and W. Huang, J. Mater. Chem. C \textbf{5}, 8734 (2017).
\bibitem{e1}M. Lu, Q. S. Yao, C. Y. Xiao, C. X. Huang and E. J. Kan,  ACS Omega \textbf{4}, 5714 (2019).

\bibitem{e2}J. Zhang, S. Jia, I. Kholmanov, L. Dong, D. Er, W. Chen,
H. Guo, Z. Jin, V. B. Shenoy, L. Shi and J. Lou, ACS Nano  \textbf{11}, 8192 (2017).





\bibitem{1}P. Hohenberg and W. Kohn, Phys. Rev. \textbf{136},
B864 (1964); W. Kohn and L. J. Sham, Phys. Rev. \textbf{140},
A1133 (1965).

\bibitem{pv1} G. Kresse, J. Non-Cryst. Solids \textbf{193}, 222 (1995).

\bibitem{pv2} G. Kresse and J. Furthm$\ddot{u}$ller, Comput. Mater. Sci. 6, \textbf{15} (1996).

\bibitem{pv3} G. Kresse and D. Joubert, Phys. Rev. B \textbf{59}, 1758 (1999).

\bibitem{pbe}J. P. Perdew, K. Burke and M. Ernzerhof, Phys. Rev. Lett. \textbf{77}, 3865 (1996).

\bibitem{u}S. L. Dudarev, G. A. Botton, S. Y. Savrasov, C. J. Humphreys and A. P. Sutton, Phys. Rev. B \textbf{57}, 1505 (1998).

\bibitem{pv5}A. Togo, F. Oba, and I. Tanaka, Phys. Rev. B \textbf{78}, 134106
(2008).

\bibitem{pv6}X. Wu, D. Vanderbilt and  D. R.  Hamann, Phys. Rev. B  \textbf{72}, 035105 (2005).


\bibitem{r1}E. Mariani and F. V. Oppen, Phys. Rev. Lett. \textbf{100}, 076801 (2008).


\bibitem{r2}J. Carrete , W. Li, L. Lindsay, D. A. Broido, L. J. Gallego and N. Mingo, Mater. Res. Lett. \textbf{4}, 204 (2016).


\bibitem{ela}R. C. Andrew, R. E. Mapasha, A. M. Ukpong and N. Chetty, Phys. Rev. B \textbf{85}, 125428 (2012).


\bibitem{v2}P. Zhao, Y. Dai, H. Wang, B. B. Huang and  Y. D. Ma, ChemPhysMater, \textbf{1}, 56 (2022).



\bibitem{v3}R. Li, J. W. Jiang, W. B. Mi  and H. L. Bai, Nanoscale  \textbf{13}, 14807 (2021).

\bibitem{q9}J. Tan, Y. H. Wang, Z. T. Wang, X. J. He, Y. L. Liu, B. Wanga, M. I. Katsnelson and  S. J.  Yuan, Nano Energy \textbf{65},  104058 (2019).

\bibitem{re-3} J. Qiu, H. Li,  X. P. Chen et al.,  J. Appl. Phys. \textbf{129}, 125109 (2021).

\bibitem{o1}A. A. M. Noor, H. J. Kim and Y. H. Shin,  Phys. Chem.
Chem. Phys. \textbf{16}, 6575 (2014).


\bibitem{o2} M. T. Ong and E. J. Reed, ACS Nano \textbf{6},  1387 (2012).
\bibitem{q7-6-1}Y. Guo, S. Zhou, Y. Z. Bai, and J. J. Zhao, Appl. Phys. Lett. \textbf{110}, 163102 (2017).


\bibitem{o3}S. D. Guo, X. S. Guo, Z. Y. Liu and Y. N. Quan,  J. Appl. Phys. \textbf{127}, 064302 (2020).

\bibitem{o4}Z. Kahraman, A. Kandemir, M. Yagmurcukardes  and H. Sahin, J. Phys. Chem. C  \textbf{123},  4549 (2019).

\bibitem{re-11}M. Yagmurcukardes and F. M. Peeters, Phys. Rev. B  \textbf{101}, 155205 (2020).





\bibitem{re-6}Z. Kahraman, A. Kandemir, M. Yagmurcukardes  and H. Sahin, J. Phys. Chem. C  \textbf{123}, 4549 (2019).


\bibitem{h10}S. Li, Q. Q. Wang, C. M. Zhang, P. Guo and S. A. Yang,  Phys. Rev. B  \textbf{104}, 085149 (2021).

\bibitem{h11} H. Hu, W. Y. Tong, Y. H. Shen, X. Wan, and C. G. Duan, npj
Comput. Mater. \textbf{6}, 129 (2020).




\end{references}
\end{document}